\documentclass[twocolumn, prl, superscriptaddress]{revtex4-1}
\usepackage{amssymb}
\usepackage{amsmath}
\usepackage{epsfig}
\usepackage{color}
\usepackage{graphics, graphicx}
\usepackage{bbold}
\usepackage{psfrag}
\usepackage{mathcomp}
\usepackage{subfigure}
\usepackage{verbatim}
\usepackage{color}
\usepackage[colorlinks,citecolor=blue]{hyperref}

\begin{document}
\title{Local manipulation of quantum magnetism in 1D ultracold Fermi gases across narrow resonances}
\author{Lei Pan}
\affiliation{Beijing National Laboratory for Condensed Matter Physics, Institute of Physics, Chinese Academy of Sciences, Beijing 100190, China}
\affiliation{School of Physical Sciences, University of Chinese Academy of Sciences, Beijing 100049, China}
\author{Xiaoling Cui}
\email{xlcui@iphy.ac.cn}
\affiliation{Beijing National Laboratory for Condensed Matter Physics, Institute of Physics, Chinese Academy of Sciences, Beijing 100190, China}
\date{\today}
\author{Shu Chen}
\email{schen@iphy.ac.cn}
\affiliation{Beijing National Laboratory for Condensed Matter Physics, Institute of Physics, Chinese Academy of Sciences, Beijing 100190, China}
\affiliation{School of Physical Sciences, University of Chinese Academy of Sciences, Beijing 100049, China}
\affiliation{Collaborative Innovation Center of Quantum Matter, Beijing, China}

\begin{abstract}
Effective range is a quantity to characterize the energy dependence in two-body scattering strength, and is widely used in cold atomic systems especially across narrow resonances. Here we show that the effective range can significantly modify the magnetic property of one-dimensional (1D) spin-$1/2$ fermions in the strongly repulsive regime.
In particular, the effective range breaks the large spin degeneracy in the hard-core limit, and induces a Heisenberg exchange term in the spin chain that is much more sensitive to the local density than that induced by the bare coupling.
With an external harmonic trap, this leads to a very rich  magnetic pattern where the anti-ferromagnetic(AFM) and ferromagnetic (FM) correlations can coexist and distribute in highly tunable  regions across the trap.
Finally, we propose to detect the range-induced magnetic order in the tunneling experiment. Our results can be directly tested in 1D Fermi gases across narrow resonance, and suggest a convenient route towards the local manipulation of quantum magnetism in cold atoms.
 \end{abstract}

\maketitle

{\it Introduction.} Energy dependent interaction is common in nature, which roots deeply in the renormalization group theory. In cold atomic systems, such energy dependence appears naturally in the Feshbach resonance which essentially relies on the energy difference of (open) atomic  and (closed) molecular channels\cite{Chin}. To describe the low-energy physics, an effective range expansion is usually introduced to incorporate the energy dependence of coupling strength $g(E)$, which reads
\begin{equation}
\frac{1}{g(E)}=\frac{1}{g(0)} + r_0 E,
\end{equation}
where $E$ is the energy of two colliding particles in the center-of-mass frame, and $g(0)$ is coupling strength at threshold energy. Here $r_0$ is the effective range, which crucially depends on the width of resonance. In particular, for narrow resonances, 
$r_0$ is typically large as to be comparable with the inter-particle distance. Previous studies have revealed interesting effects of finite range in cold atoms systems, including the generation of stronger interaction effects\cite{Ho, Ohara, Zhai, Cui, Guan}, the modification of Fermi superfluids\cite{Gurarie,Nishida}, the sub-leading high-momentum tail\cite{Platter, Yu_p, Zhou, Cui2,Joseph,Yin}, and the stabilization of repulsive polaron\cite{Grimm} and p-wave system\cite{Pan}.

In this work, we reveal the significant effect of effective range to the quantum magnetism of spin-$1/2$ Fermi gases. Here we take the 1D Fermi gas in strongly repulsive regime, which features an impenetrable nature and thus supports a hidden ``lattice" structure. In this regime, the system is well described by an effective Heisenberg spin-chain Hamiltonian which exhibits exotic quantum magnetic properties and provides additional insights into the simulation of quantum
magnetism  without lattice\cite{Zinner2,Santos,Pu,Hu,Volosniev,Massignan,Pu2, Hu2,Deuretzbacher,Deuretzbacher2,Pan2,Levinsen,Yang-Cui,Cui_p} and such spin-chain systems were realized experimentally recently\cite{Jochim_chain}.
Previous studies based on zero-range interactions have shown that the system can host either an anti-ferromagnetic (AFM) or a ferromagnetic (FM) spin correlation, depending on the sign of coupling strength or other small perturbations. Moreover, at the infinite coupling (hard-core) limit, spin and charge are fully decoupled and the system exhibits huge spin degeneracy, at which point the FM transition is predicted\cite{Cui-Ho}. Here, we will show that the inclusion of a finite effective range can qualitatively change above conclusions and brings much richer magnetic structures to the system. 

\begin{figure}[t]
\includegraphics[width=8cm]{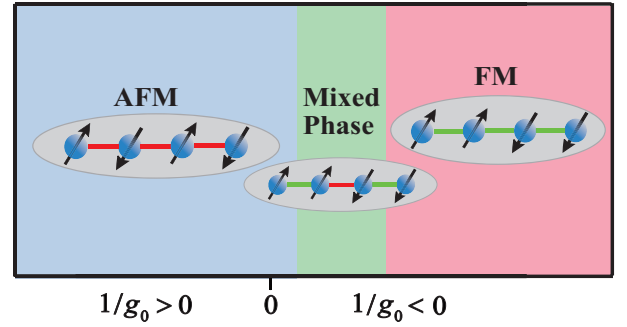}
\caption{(Color online).
Schematic diagram of magnetic order of a spin-$1/2$ fermion system with a finite-range ($r_0\neq0$) interaction in a harmonic trap. As we tune the interaction ($1/g_0$) from the repulsive to the attractive side, the system adiabatically goes from a AFM-
correlated spin chain to FM-AFM-FM mixed and finally to
fully FM-correlated chain, which is in sharp contrast to the zero-range case with either AFM phase ($1/g_0>0$) or FM phase($1/g_0<0$). The red and green bonds represent the AFM and FM couplings between spins, respectively.} \label{fig_schematic}
\end{figure}

Our results can be summarized in Fig.\ref{fig_schematic}. In the presence of a finite range $r_0$, as tuning the inverse coupling $1/g_0$ from the repulsive  to attractive side, the system adiabatically goes from a AFM-correlated spin chain to FM-AFM-FM mixed and finally to fully FM-correlated chain. The spatially modulated magnetic correlation is due to the range-modified Heisenberg coupling in an effective spin chain model, which is more sensitive to the local density as compared to that induced by bare coupling (see Eq.\ref{J}).
Our results demonstrate an adiabatic formation of FM domains from the AFM state, which has not been achieved up to date. We further propose to verify these magnetic properties in the tunneling experiment of tilted harmonic potential.


{\it Range-modified effective spin chain.} We begin with deriving an effective spin chain model for strongly repulsive spin-$1/2$ fermions($\uparrow,\downarrow$) in the presence of a finite range. The original Hamiltonian is $H=H^{(0)}+U$,  (here $\hbar=1$)
\begin{eqnarray}
H^{(0)}&=&\sum_i \left( -\frac{1}{2m} \frac{\partial^2}{\partial x_i^2} + \frac{1}{2} m\omega_{ho}^2 x_i^2 \right), \label{H0}\\
U&=&   \sum_{i,j} g_{ij}\delta(x_{i\uparrow}-x_{j\downarrow}). \label{Us} \end{eqnarray}
Here $g_{ij}$ follows the effective range expansion: $1/g_{ij}=1/g_0+r_0 E_{ij}$, with $g_0$ the bare coupling constant and $E_{ij}$ the relative energy of two colliding particles $x_{i\uparrow}$ and $x_{j\downarrow}$.   The present study will focus on the near resonance regime with large $g_0$ and small $r_0$. Here the confinement length is defined as $a_{ho}=(m\omega_{ho})^{-1/2}$.

To highlight the range effect, let us first consider the case of $g_0=\infty$. In this case, without the range ($r_0= 0$) the collision of atoms is forbidden due to hard-core interaction and the spins can distribute in an arbitrary order in coordinate space, giving the large spin degeneracy. When turn on the range ($r_0\neq 0$), however, the atoms only experience hard-core interaction at zero relative energy($E_{rel}=0$) but not at finite $E_{rel}$, and the finite-$E_{rel}$ interaction causes a super-exchange of spins at neighboring orders in the coordinate space, giving rise to an effective spin chain model. Following the standard procedure, we obtain the effective Heisenberg spin-chain solely induced by $r_0$:
\begin{equation}
H_{\rm eff}^r= r_0\sum_l J^r_l  \left({\bf s}_l\cdot {\bf s}_{l+1} -\frac{1}{4}\right), \label{H_r}
\end{equation}
here $l$ is the order index of particles in coordinate space, and the Heisenberg coupling is
\begin{eqnarray}
J^r_l&=&\frac{2N!}{m^2}  \int d{\bf x}  E_{ij}\Big|\frac{\partial D}{\partial x_{ij}}|_{x_{ij}=0} \Big|^2 \theta(\cdots<x_i=x_j<\cdots),       \label{J_r}
\end{eqnarray}
where $x_{ij}=x_i-x_j$; $D(\{x_i\})$ is the Slater determinant of N fermions occupying the lowest N-level of $H^{(0)}$, and $E_{ij}$ is the relative collision energy of two particles ($x_i,\ x_j$) in the $D(\{x_i\})$; in the $\theta$-function $x_i(=x_j)$ is with order index $l$.

\begin{figure}[t]
\includegraphics[width=8.0cm]{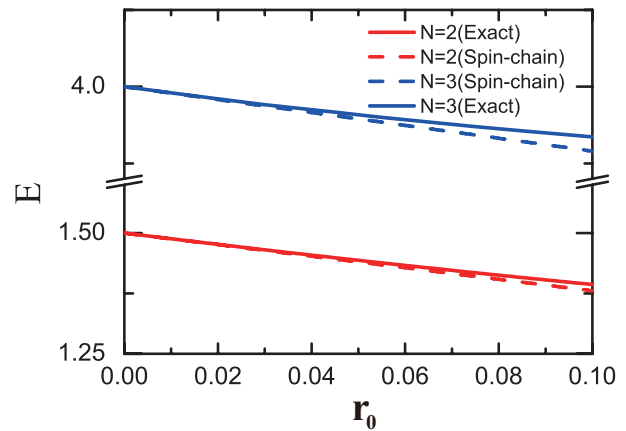}
\caption{(Color online). Energy spectrum for 1D trapped $\uparrow\downarrow$ and $\uparrow\uparrow\downarrow$ systems as a function of effective range $r_0$ at 1D resonance ($1/g_0=0$). The linear fit is based on the effective spin-chain model (\ref{H_r}).  Here the energy and length units are $\omega_{ho}$ and $a_{ho}$ respectively.} \label{fig_few}
\end{figure}

To verify the range-induced spin-chain model, we have exactly solved the two-body ($\uparrow\downarrow$) and three-body ($\uparrow\uparrow\downarrow$) problems in a trapped system with tunable range. 
In Fig.\ref{fig_few}, we plot the obtained energy spectra of these systems as a function of $r_0$ at $1/g_0=0$, in comparison with the  linear fit from the model $H_{\rm eff}^r$. We can see that the effective model can well predict the real spectra of both systems for small $r_0(\le 0.1a_{ho})$.  This validates the effective spin-chain model used for larger systems.

In combination with the spin-chain model for small $1/g_0$ \cite{Santos,Pu,Hu,Volosniev,Massignan,Pu2, Hu2,Deuretzbacher,Levinsen,Cui_p,Yang-Cui}, we write the final effective model in the limit of large $g_0$ and small $r_0$ as:
\begin{equation}
H_{\rm eff}= \sum_l \left(\frac{1}{g_0}J_l^g + r_0J^r_l\right)  \left({\bf s}_l\cdot {\bf s}_{l+1} -\frac{1}{4}\right), \label{H_tot}
\end{equation}
with $J_l^g$:
\begin{eqnarray}
J^g_l&=&\frac{2N!}{m^2} \int d{\bf x}  \Big|\frac{\partial D}{\partial x_{ij}}|_{x_{ij}=0} \Big|^2 \theta(\cdots<x_i=x_j<\cdots).       \label{J_g}
\end{eqnarray}
Physically, both the bare coupling ($g_0$) and effective range ($r_0$)  produce the same isotropic Heisenberg term is because both of them take effect in the spin-singlet interaction channel, thus the effective model is determined by the same spin-projection operator \cite{Yang-Cui}.

\begin{figure}[t]
\includegraphics[width=8.5cm]{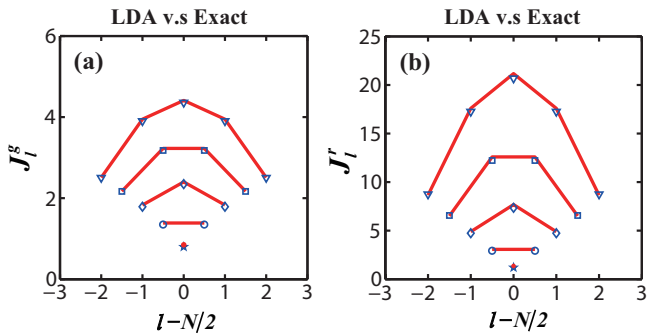}
\caption{(Color online). Heisenberg couplings $J^g_l$ (a) and $J^r_l$ (b) (in the units of $\frac{1}{m^2a_{ho}^3}$ and $\frac{1}{m^3a_{ho}^5}$ respectively) in a harmonic trap. The star, circle, diamond, square, and triangular points are exact solutions of Eqs.(\ref{Jg},\ref{Jr}) for total number $N=2,3,4,5,6$. The red lines are from analytical expressions (\ref{J}) together with Thomas-Fermi density (see text).  } \label{fig_J}
\end{figure}

{\it Heisenberg couplings.} Before studying the quantum magnetism, we first examine the density dependence of $J_l^g,\ J_l^r$ for a homogeneous large system, and the trapped case can be deduced from the local density approximation(LDA).  Previously, $J_l^g$ was shown to depend on the cubic density $(\sim n^3)$\cite{Santos,Marchukov}, by extrapolating the nearest-neighboring exchange coupling in Hubbard model to continuum\cite{Ogata,Matveev}. Here we point out an alternative way to derive $J_l^g$ and $J_l^r$ from Eqs.(\ref{J_g}, \ref{J_r}) through the momentum averaging below the Fermi sea:
\begin{eqnarray}
J^g&=& \frac{2n}{m^2} \Big\langle  \left( \frac{k_1-k_2}{2} \right)^2 \Big\rangle, \label{Jg} \\
J^r&=& \frac{2n}{m^3} \Big\langle  \left( \frac{k_1-k_2}{2} \right)^4 \Big\rangle, \label{Jr}
\end{eqnarray}
here $\langle  F(k_1,k_2) \rangle\equiv \int\int F(k_1,k_2) dk_1dk_2/\int\int dk_1dk_2$, and the integration is for $k_1,k_2\in[-k_F,k_F]$, with  $k_F$ the Fermi momentum determined by the  density $n=k_F^3/(6\pi^2)$. The essence of Eqs.(\ref{Jg},\ref{Jr}) is to reformulate the many-fold integration into the combination of the local density and a pair-averaged function in terms of the relative momentum of two particles within the Fermi sea. The procedure leads to
\begin{equation}
J^g=\frac{2\pi^2n^3}{3m^2},\ \ \ \ J^r=\frac{4\pi^4n^5}{15m^3}. \label{J}
\end{equation}
Remarkably, here the range-induced coupling $J^r$ has a much more sensitive dependence on the local density than $J^g$, which we will show below to significantly affect the quantum magnetism in the trapped system.
In Fig.\ref{fig_J}, we show Eq.\ref{J} can well reproduce the exact solutions of $J^g_l,\ J^r_l$ from Eqs.(\ref{Jg},\ref{Jr}) for trapped systems up to $N=6$, here for the local density we have used  Thomas-Fermi approximation $n_l\rightarrow n(\bar{x}_l)=\frac{1}{\pi}\sqrt{2m(N\omega_T-\frac{1}{2} m\omega_T^2 \bar{x}_l^2)}$, with $\bar{x}_l=\langle (x_l+x_{l+1})/2 \rangle$.

To this end we can rewrite the spin-chain Hamiltonian (\ref{H_tot}) as $H_{\rm eff}=\sum_l J^{\rm eff}_l \left({\bf s}_l\cdot {\bf s}_{l+1} -\frac{1}{4}\right)$, where the effective coupling depends on the coupling, range, and local density $n_l$:
\begin{equation}
J^{\rm eff}_l= \frac{2\pi^2n^3_l}{3m^2} \left(\frac{1}{g_0} + \frac{2\pi^2}{5m} n_l^2r_0 \right) .\label{J_eff}
\end{equation}

{\it Spatially modulated quantum magnetism.} From the expression of $J^{\rm eff}_l$, we can see that its sign can be effectively tuned by local density $n_l$, distinct from the zero-range case where the magnetic property is solely determined by the sign of coupling strength. This immediately leads to two effects for trapped system with inhomogeneous density.
First, the system is no longer described by a single coupling strength, and there is no exact  hard-core limit  with large spin degeneracy.
Secondly, given $r_0>0$ and in the regime of $1/g_0<0$, particles at different regions inside the trap may experience different signs of $J^{\rm eff}$, which means that the AFM and FM magnetic correlation can coexist in the system, i.e., the quantum magnetism can be locally manipulated.


\begin{figure}[h]
\includegraphics[height=5cm]{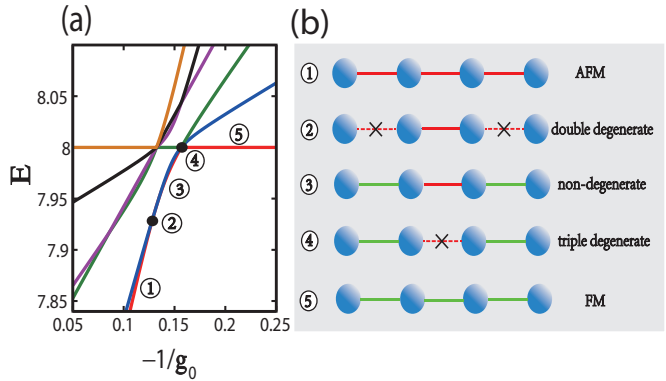}
\caption{(Color online). (a) Energy spectrum of four fermions ($N_{\uparrow}=N_{\downarrow}=2$) as a function of $-1/g_0$(in the unit of $ma_{ho}$) in a trapped system. Here we take $r_0=0.05a_{ho}$. (b) Schematics of AFM (red lines) and FM (green lines) magnetic correlation in the chain at different coupling strengths as marked in (a). The dashed lines with crossings refers to the zero effective coupling $J^{\rm eff}_l=0$. } \label{fig_spectrum}
\end{figure}


In Fig.\ref{fig_spectrum}(a), we show the energy spectrum of four fermions ($N_{\uparrow}=N_{\downarrow}=2$) by solving the spin-chain Hamiltonian (\ref{H_tot}) at a given $r_0=0.05a_{ho}$. Totally there are six energy levels, two with total spin $S=0$, three with $S=1$ and one with $S=2$, similar to the zero-range case\cite{Cui}.
With a finite range, the six levels no longer cross each other at $1/g_0=0$, while the ground state transition (from $S=0$ to $S=2$) moves to the negative coupling side at $g_c<0$.

Now we analyze the adiabatic change of the ground state (with $S=0$) before the transition (in the regime $-1/g_0<-1/g_c$). In the positive coupling side of resonance, the ground state holds the AFM correlations in the spin chain given by all positive $J_l^{\rm eff}$(marked as \textcircled{1} in Fig.\ref{fig_spectrum}(a,b)). As increasing $-1/g_0$ across resonance to negative coupling side, two states becomes degenerate at point \textcircled{2}, where the Heisenberg coupling at the edges of the chain touches zero($J^{\rm eff}_1=J^{\rm eff}_3=0$) and the edge particles are decoupled from the chain, giving two degenerate spin states ($S=0$ and $S=1$). Further increasing $1/g_0$ beyond this point, the edge spins have FM correlation with negative coupling, while the center are AFM correlated with positive coupling (see \textcircled{3}). When reaching \textcircled{4},  the center coupling becomes zero ($J^{\rm eff}_2=0)$, and the system is divided into two independent magnetic domains with FM correlation. At this point, three spin states are degenerate, corresponding to two FM(triplet) pairs forming total spin $S=0,1,2$. Beyond this point, the ground state of the system changes to $S=2$ FM state, and all sites are with FM correlations (\textcircled{5}).

\begin{figure}[h]
\includegraphics[height=6.5cm]{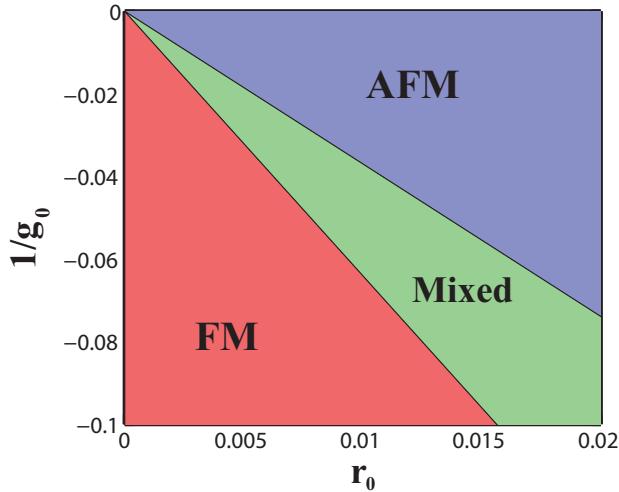}
\caption{(Color online). Phase diagram of magnetic order in terms of the bare coupling and effective range. Here $N_{\uparrow}=N_{\downarrow}=4$. The units of $1/g_0$ and $r_0$ are $ma_{ho}$ and $a_{ho}$ respectively.} \label{fig_phase}
\end{figure}

Above picture can be generalized to an arbitrary number of particles. Take the spin-balanced case $N_{\uparrow}=N_{\downarrow}=N/2$ for example, as increasing $-1/g_0$ the ground state of the system(with $S=0$) is expected to cross a sequence of degeneracies with other spin states until the transition to FM ($S=N/2$) state. The first degeneracy occurs at $1/g_0=-\frac{2\pi^2}{5m}n_1^2r_0$, when $J^{\rm eff}_1=J^{\rm eff}_{N-1}=0$ and two edge spins are separated from the system giving a two-fold degeneracy, see the upper line in Fig.\ref{fig_phase}. Increasing $-1/g_0$ beyond this point, the Heisenberg couplings transit from pure AFM type to FM-AFM-FM mixed type. In this mixed phase, the trap center shows the AFM correlation (with $J^{\rm eff}>0$) because of higher density, while the trap edge shows FM correlation with $J^{\rm eff}>0$ because of lower density. In this regime, a $m+1$-fold ($1\le m\le N/2)$ degeneracy occurs at $1/g_0=-\frac{2\pi^2}{5m}n_m^2r_0$ when $J^{\rm eff}_m=J^{\rm eff}_{N-m}=0$. Continuously increasing $-1/g_0$, the regions with FM correlations becomes enlarged while AFM correlation becomes reduced, giving the increasing $\langle {\bf S}_L^2{\bf S}_R^2 \rangle$,  where $S_{L/R}$ is the total spin of left/right part of the trapped system.  The ground state transition to FM state happens exactly at the $N/2+1$ fold degeneracy point(see lower phase boundary in Fig.\ref{fig_phase}), when $1/g_0=-\frac{2\pi^2}{5m}n(0)^2r_0$ with $n(0)$ the density at trap center. At this point $\sqrt{\langle {\bf S}_L^2{\bf S}_R^2 \rangle}$ reaches the maximum value $N/4(N/4+1)$, suggesting two large and separated FM domains formed at the left and right regions of the trap.

{\it Tunneling experiment}. Now we come to the experimental detection of the range-induced magnetic orders, using the tunneling techniques as established in experiments\cite{Jochim_chain}. By varying the magnetic field gradient and tilting the potential barrier, one can control the number of atoms tunneling out of the trap and probe the spin structure. Here we propose the measure the possibility of having all spin-$\downarrow$ atoms tunneling from the right side  of the trap,  which is given by the weight of the full spin separated configuration ($|\uparrow\uparrow...\downarrow\downarrow...\rangle$) in the wave function, as denoted by $P_{\downarrow}$ in following discussions.

\begin{figure}[h]
\includegraphics[height=4.5cm]{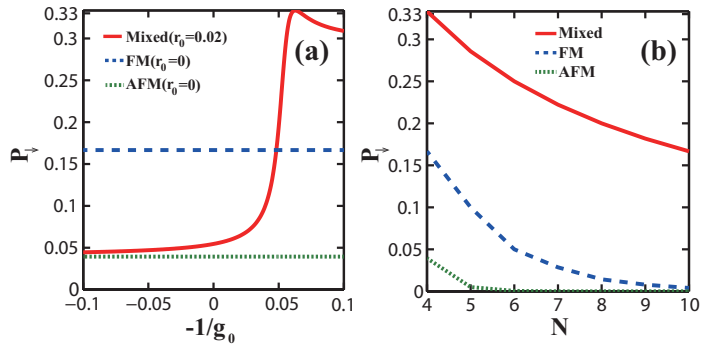}
\caption{(Color online). (a) Probabilities of all spin-$\downarrow$ atoms tunneling from the right side of the titled trap, denoted by $P_{\downarrow}$. Red solid line is for system with range $r_0=0.02a_{ho}$, by adiabatically following the ground state in $g_0>0$ side to  $g_0<0$ side. The cross point marks the location of FM transition, when $P_{\downarrow}$ shows a maximum. Blue dashed and green dotted lines are for zero-range systems adiabatically following the AFM ground state($S=0$) and the FM state($S=N/2$). Here $N_{\uparrow}=N_{\downarrow}=2$, the unit of $1/g_0$ is $ma_{ho}$. (b) The maximum of $P_{\downarrow}$ as a function of particle number $N$ for spin-balanced fermions. Red, blue and green lines are the same as in (a).  } \label{fig_tunnel}
\end{figure}

In Fig.\ref{fig_tunnel} (a), we show $P_{\downarrow}$ as a function of $-1/g_0$ for $N_{\uparrow}=N_{\downarrow}=2$ system, by adiabatically following certain state from the  $g_0>0$ side. Without range, we see that $P_{\downarrow}$ is very small ($\sim 4\%$) if following the AFM ground state, and can be as large as $16.7\%$ if following FM state, consistent with the experiment \cite{Jochim_chain}. Turning on the range and following the ground state in $g_0>0$ side, $P_{\downarrow}$ is no longer a constant but varies sensitively with $-1/g_0$, suggesting the significant change of magnetic structures/correlations in the trap. In particular, we see that $P_{\downarrow}$ reaches a maximum near the $N/2+1$-fold degeneracy point, where the effective coupling at the trap center touches zero and the system is composed of two FM domains (each with spin $S=N/4$). The maximum value can be then deduced by expanding the $S=0$ state by two spins with $S=N/4$:
\begin{equation}
|S=0,S_z=0\rangle=\sum_{m=-\frac{N}{4}}^{\frac{N}{4}}C_m^{(N)}|S_1=\frac{N}{4},m\rangle_{\frac{N}{2}}|S_2=\frac{N}{4},-m\rangle_{\frac{N}{2}},
\label{FM_point}
\end{equation}
then $P_\downarrow$ is exactly given by the Clebsch-Gordor coefficients as:
\begin{equation}
P_\downarrow=
\Big|C^{(N)}_\frac{N}{4}\Big|^2=\frac{1}{N/2+1}. \label{P_max}
\end{equation}
In Fig.\ref{fig_tunnel}(b), we have verified this analytic result by numerically calculations from the spin-chain model. Remarkably, Eq.\ref{P_max} produces a much larger $P_{\downarrow}$ in comparison to the  FM state, where $P_{\downarrow}=\frac{\frac{N}{2}!\frac{N}{2}!}{N!}\simeq\frac{\sqrt{\frac{\pi N}{2}}}{2^N}$ is exponentially small. This reflect the distinct magnetic structure of a pure FM state and a composition of two FM domains. This can serve as experimental evidence to identify the spatially-modulated quantum magnetism due to the finite range effect.

{\it Final remark.} Our results reveal the significant effect of a finite effective range in the strong coupling regime of 1D trapped spin-$1/2$ fermions. The sensitive density-dependence of Heisenberg coupling induced by the finite range suggests a convenient route towards the local manipulation of quantum magnetism. In particular, by engineering the density distribution of cold atomic gases through the laser potentials, one may get access to an arbitrary configuration of local Heisenberg coupling in the coordinate space, and thus an arbitrary type of magnetic order may be achieved. This concept can be generalized to higher spins and other composition of atomic mixtures in the 1D strong coupling regime.

{\it Acknowledgment.}  The work is supported by the National Key Research and Development Program of China (2018YFA0307600, 2016YFA0300603), and the National Natural Science Foundation of China (No.11622436, No.11425419, No.11421092, No.11534014).

\end{document}